# Implementing the Typed Graph Data Model Using Relational Database Technology

Malcolm Crowe
Emeritus Professor, Computing Science
University of the West of Scotland
Paisley, United Kingdom
Email: Malcolm.Crowe@uws.ac.uk

Fritz Laux
Emeritus Professor, Business Computing
Reutlingen University
Reutlingen, Germany
Email: Fritz.Laux@reutlingen-university.de

*Abstract*—Recent standardization work for database languages has reflected the growing use of typed graph models (TGM) in application development. Such data models are frequently only used early in the design process, and not reflected directly in underlying physical database. In previous work, we have added support to a relational database management system (RDBMS) with role-based structures to ensure that relevant data models are not separately declared in each application but are an important part of the database implementation. In this work, we implement this approach for the TGM: the resulting database implementation is novel in retaining the best features of the graph-based and relational database technologies.

## I. INTRODUCTION

The work in this paper was signaled in a conference presentation [1] in early 2023 and reflects ongoing work in the standardization community to create standards for graph databases. This has already led to the adoption of a new chapter in the International Standards Organization (ISO) standard 9075 [2] for property graph queries, and a draft international standard (DIS) on Graph Query Language (GQL) is now expected in early 2024.

Many data models assist in the development of software, such as the Unified Modeling Language (UML) [3][4], entity frameworks, and persistence architectures. During such early conceptual model building, incremental and interactive exploration can be helpful [5] as fully automated integration tools may combine things in an inappropriate way, and the use of data types [6] can help to ensure that semantic information is included not merely in the model, but also in the final database. In this short paper we report on such an implementation of the Typed Graph Model (TGM), using metadata in a relational database management system (RDBMS) [7], partly inspired by recent developments in the PostgreSQL community [8]. Some recent database management systems (DBMS) have included metadata in the relational model to form a bridge with the physical database, so that the data model can be enforced across all applications for a single database. In this work, we provide a mechanism for integrating the graphical data model in the physical RDBMS.

As with the original relational model, the TGM has a rigorous mathematical foundation as an instance of a Graph Schema.

The plan of this paper is to review the TGM in Section II, and discuss the implementation details in Section III. Section IV presents an illustrative example, and Section V provides some conclusions.

## II. THE TYPED GRAPH MODEL AND INFORMATION INTEGRATION

We will construct a TGM for a database by declaring instances of nodes and edges as an alternative to specifying tables of nodes and edges.

### A. Typed Graphs Formalism

In this section we review the informal definition of the TGM from [2], using small letters for elements (nodes, edges, data types, etc.) and capital letters for sets of elements. Sets of sets are printed as bold capital letters. A typical example would be $n \in N \in \mathbf{N} \subseteq \wp(N)$, where $N$ is any set and $\wp(N)$ is the power-set of $N$.

Let $T$ denote a set of simple or structured (complex) data types. A data type $t:=(l,d) \in T$ has a name $l$ and a definition $d$. Examples of simple (predefined) types are ($int,\mathbb{Z}$), ($char,ASCII$), ($\%,[0..100]$), etc. It is also possible to define complex data types like an order line (*OrderLine*, (*posNo*, *partNo*, *partDescription*, *quantity*)). The components need to be identified in $T$, e. g., ($posNo,int>0$). Recursion is allowed as long as the defined structure has a finite number of components.

The UML-notation was chosen as graphical representation for nodes and include the properties as attributes including their data types. Labels are written in the top compartment of the UML-class. Edges of the TGS are represented by UML associations. For the label and properties of an edge we use the UML-association class, which has the same rendering as an ordinary class, but its existence depends on an association (edge), which is indicated by a dotted line from the association class to the edge. This not only allows to label an edge but to define user defined edge types. The correspondence between the UML notation and the TGS definition is shown in Table I.

**Definition 1 (Typed Graph Schema, TGS)** *A typed graph schema is a tuple* $TGS=(N_S,E_S,\wp,T,\tau,C)$
*where:*

- $N_S$ is the set of named (labeled) objects (nodes) n with properties of data type $t:=(l,d)\in T$, where l is the label and d the data type definition.
- $E_S$ is the set of named (labeled) edges e with a structured property $p:=(l,d)\in T$, where l is the label and d the data type definition.
- $\varrho$ is a function that associates each edge e to a pair of object sets (O,A), i. e., $\varrho(e):=(O_e,A_e)$ with $O_e, A_e \in \wp(N_S)$. $O_e$ is called the tail and $A_e$ is called the head of an edge e.
- $\tau$ is a function that assigns for each node n of an edge e a pair of positive integers $(i_n,k_n)$, i. e., $\tau_e(n):=(i_n,k_n)$ with $i_n \in N_0$ and $k_n \in N$. The function $\tau$ defines the min-max multiplicity of an edge connection. If the min-value $i_n$ is 0 then the connection is optional.
- C is a set of integrity constraints, which the graph database must obey.

The notation for defining data types T, which are used for node types $N_S$ and edge types $E_S$, can be freely chosen: and in this implementation SQL will be used for identifiers and expressions, together with a strongly typed relational database engine. The integrity constraints C restrict the model beyond the structural limitations of the multiplicity τ of edge connections. Typical constraints in C are semantic restrictions of the content of an instance graph. For instance, in an order processing graph-database a constraint should require that an "order"-node o should have at least one "order-detail" node ol connected by an edge labelled "belongs_to" (see example order GDB in Table II.)

**Definition 2 (Typed Graph Model)** *A typed graph Model is a tuple TGM=(N,E,TGS,φ) where:*

- *N is the set of named (labeled) nodes n with data types from $N_S$ of schema TGS.*
- *E is the set of named (labeled) edges e with properties of types from $E_S$ of schema TGS.*
- *TGS is a typed graph schema as defined above..*
- *φ is a homomorphism that maps each node n and edge e of TGM to the corresponding type element of TGS, formally:*

$$\varphi: TGM \rightarrow TGS$$
$$n \mapsto \varphi(n) := n_S (\in N_S)$$
$$e \mapsto \varphi(e) := e_S (\in E_S)$$

The fact that φ maps each element (node or edge) to exactly one data type implies that each element of the graph model has a well-defined data type. The homomorphism is structure preserving. This means that the cardinality of the edge types is enforced, too. In our Pyrrho implementation, the declaration of nodes and edge of the TGM develops the associated TGS incrementally including the development of the implied type system T. Data type and constraint checking is applied for all nodes and edges before any insert, update, or delete action can be committed.

*B. The Data Integration Process*

The full benefit of information integration requires the integration of source data with their full semantics. We believe a key success factor is to model the sources and target information as accurately as possible. The expressive power and flexibility of the TGM allows precise description of the meta-data of the sources and target in the same model, which simplifies the matching and mapping of the sources to the target. The tasks of the data integration process are:

1) model sources as TGS $S_i (i = 1, 2, ..., n)$
2) model target schema *T* as TGS *G*
3) match and map sources $S_i$ with TGS *G*
4) check and improve quality
5) convert TGS G back to T again

Steps 3 and 4 can occur together in an interactive process once the basic model has been outlined. Such a process is crucial for Enterprise Information Integration (EII) and other data integration projects, which demand highly accurate information quality, which can be further improved with the use of different mappings.

To start the process, it may be necessary to collect structure and type information from a data expert or from additional information. Where sources are databases, the rigid structures provide a good starting point. Otherwise, the relevant data must first be identified together with its meta-data if available. This includes coding and names for the data items. The measure units and other meta-data provided by the data owner are used to adjust all measures to the same scale. The paper of Laux [6] gives some examples how to transform relational, object oriented, and XML-schemata into a TGS.

If the source is unstructured or semi-structured, e.g., documents or XML/HTML data, concepts and mechanisms from Information Retrieval (IR) and statistical analysis may help to identify some implicit structure or identify outliers and other susceptible data. If the data are self-describing (JSON, key-value pairs, or XML) linguistic matching can be applied with additional help from a thesaurus or ontology. Nevertheless, it is advisable to validate the matching with instance data or an information expert.

The use of hyper-nodes $n \in N_S$ and hyper-edges $e \in E_S$ instead of simple nodes resp. edges allow to group nodes and edges to higher abstracted complex model aggregates. This is particularly useful to keep large models clearly represented and manageable. Each sub-graph can be rendered as a hyper-node. If the division is disjoint these hyper-nodes are connected via hyper-edges forming a higher abstraction level schema.

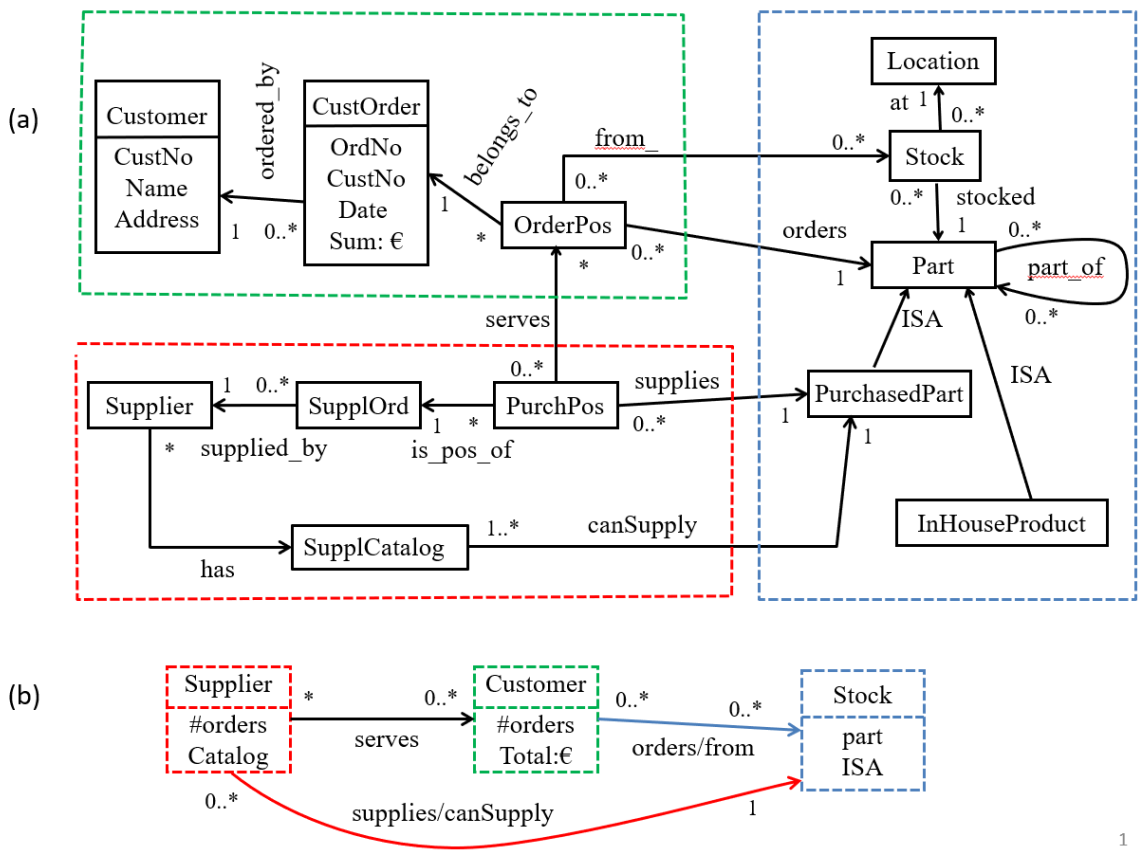

Figure 1. Example TGM of a commercial enterprise showing two levels of detail

We present two possible TGS abstraction levels for a single enterprise in UML notation in Figure 1. The dashed green line in part (a) encompasses the Customer data, comprising Customer master data, customers' orders, and order positions. In part (b) this information is concentrated in one node type named "Customer". The Supplier data side (red dashed line) is modelled in the same manner. The stock management data show in the detailed part (a) the bill of material (BOM) which is modelled as a recursive edge "Part of" on the parts node. This is no longer explicitly visible in the aggregated part (b). This hidden information should be part of the now complex property "part" of the Stock node.

This little example demonstrates already the flexibility of the model in terms of detail and abstraction. We discuss this example in some detail in Section IV.

### III. IMPLEMENTATION IN THE RELATIONAL DATABASE SCHEMA

The implementation of a typed graph modelling system can build on the user-defined type mechanism of an RDBMS. Node and edge types should have special columns: for node types, there is an automatic primary key with default name ID, and edge types also automatic foreign keys for their source and destination nodes, that are referred to here by their default names LEAVING and ARRIVING, and these should have automated support from the RDBMS. It should be possible to convert between standard types and node/edge types and rearrange subtype relationships. These tables can be equipped with indexes, constraints, and triggers in the normal ways.

Then, if every node type or edge type corresponds to a single base table containing the instances of that type, one way to build a graph is to insert rows in these tables. But a satisfactory implementation needs to simplify the tasks of graph definition and searching. Most implementations add CREATE and MATCH statements, which we describe next, and indicate how they can be implemented in the RDBMS.

#### A. Graph-oriented Syntax Added to SQL

The typical syntax for CREATE sketches nodes and edges using additional arrow-like tokens, for example:

```
[CREATE    (:Person    {name:'Fred    Smith'})<-
[:Child]-(a:Person {name:'Peter Smith'}),
(a)-[:Child]->(b:Person {name:'Mary Smith'})
-[:Child]->(:Person {name:'Lee Smith'}),
(b)-[:Child]->(:Person {name:'Bill Smith'})]
```

Without any further declarations, this builds a graph with nodes for Person and edges for Child, as in Figure 2.

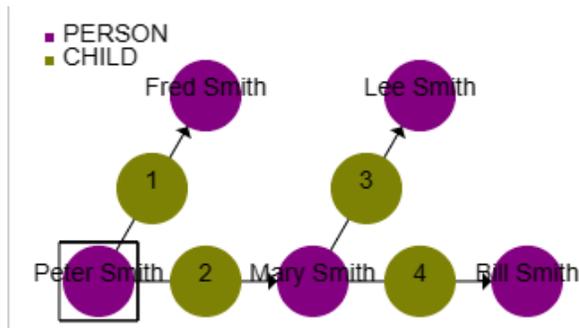

Figure 2. Browser output for web address
http://localhost:8180/ps/PS/PERSON/NAME='Peter Smith'?NODE

There is already a standard abstract syntax [9][10], that can be represented as:

CREATE Graph {',' Graph} [THEN Statement].

Graph = Node Path {',' Node Path } .
Path = { Edge Node } .
Node = '(' GraphItem ')' .
Edge = '-[' GraphItem ']->' | '<-[' GraphItem ']-'.
GraphItem = [id | *Node*_Value] [GraphLabel] [ Document ] .
GraphLabel = ':' (id | *Label*_Value) [GraphLabel] .

In this syntax, the strings enclosed in single quotes are tokens, including several new token types for the TGM. In corresponding source input, unquoted strings are used for case-insensitive identifiers and double quoted strings for case-sensitive identifiers, possibly containing other Unicode characters. As usual in SQL, string constants in input will be single quoted, and doc is a JSON-like structure providing a set of properties and value expressions, possibly including metadata definitions for ranges and multiplicity.

Nodes and edges and new node types and edge types can be introduced with this syntax. The database engine constructs a base table for each distinct label, with columns sufficient to represent the associated properties. These database base tables for node types (or edge types) contain a single row for each node (resp. edge) including node references. They can be equipped with indexes, constraints, and triggers in the normal ways.

To the normal SQL DML, we add the syntax for the MATCH query, which has a similar syntax, except that it may contain unbound identifiers for nodes and edges, their labels and/or their properties.

MatchStatement = MATCH Match {',' Match} [WhereClause] [Statement] [THEN Statements END].
Match = (MatchMode [id '='] MatchNode) {'|' Match}.

The first part of the MATCH clause has an optional MatchMode (see below) and one or more graph expressions, which in simple cases appear to have the same form as in the CREATE statement.
MatchNode = '(' MatchItem ')' {(MatchEdge|MatchPath) MatchNode}.
MatchEdge = '-[' MatchItem '->' | '<-' MatchItem ']-' .
MatchItem = [id | *Node*_Value] [GraphLabel] [ Document | WhereClause ] .

In all cases, the execution of the MATCH proceeds directly on the tables, without needing auxiliary SQL statements. The MATCH algorithm proceeds along the node expressions, matching more and more of its nodes and edges with those in the database by assigning values to the unbound identifiers. If we cannot progress to the next part of the MATCH clause, we backtrack by undoing the last binding and taking an alternative value. If the processing reaches the end of the MATCH statement, the set of bindings contributes a row in the default result, subject to the optional WHERE condition.

In this way, the MATCH statement can be used (a) as in Prolog, to verify that a particular graph fragment exists in the database, (b) to display the bindings resulting from the process of matching a set of fragments with the database, (c) to display a set of values computed from such a list of bindings, or (d) to perform a sequence of actions for each binding found. In case (d) no results are displayed, as the MATCH statement has been employed for its side effects. These could include further CREATE, MATCH or other SQL statements, or assignment statements updating fields referenced in the current bindings.

Following the forthcoming GQL standard, repeating patterns are supported by the MATCH statement (see [9]):

MatchPath = '[' Match ']' MatchQuantifier .
MatchQuantifier = '?' | '*' | '+' | '{' int , [int] '}' .
MatchMode = [TRAIL|ACYCLIC| SIMPLE] [SHORTEST |ALL|ANY] .

The MatchMode controls how repetitions of path patterns are managed in the graph matching mechanism. A MatchPath creates lists of values of bound identifiers in its Match. By default, binding rows that have already occurred in the match are ignored, and paths that have already been listed in a quantified graph are not followed. The MatchMode modifies this default behaviour: TRAIL omits paths where an edge occurs more than once, ACYCLIC omits paths where a node occurs more than once, SIMPLE looks for a simple cycle. The last three options apply to MatchStatements that do not use the comma operator, and select the shortest match, all matches or an arbitrary match.

The implementation of the matching algorithm uses continuations to control the backtracking behavior. Continuations are constructed as the match proceeds and represent the rest of the matching expression.

The MATCH statement can be used in two ways. The first is make the dependent Statement a RETURN statement that contributes a row to a result set for each successful binding of the unbound identifiers in the MATCH, for example,

```
E:\PyrrhoDB70\Pyrrho>pyrrhocmd ps
SQL> [CREATE (a:Person {name:'Fred Smith'})<-[:Child]-(b:Person {name:'Peter Smith'}),
> (a)-[:Child]->(c:Person {name:'Mary Smith'})
> -[:Child]->(:Person {name:'Lee Smith'}),
> (c)-[:Child]->(:Person {name:'Bill Smith'})]
SQL> MATCH ({name:'Peter Smith'}) [()-[:Child]->()]+ (x) RETURN x.name
|----------|
|NAME      |
|----------|
|Lee Smith |
|Bill Smith|
|Mary Smith|
|Fred Smith|
|----------|
SQL> MATCH ({name:'Peter Smith'}) [(p)-[:Child]->()]+ ({name:x})
|-----------------------------------------------------------------------------------------|----------|
|P                                                                                        |X         |
|-----------------------------------------------------------------------------------------|----------|
|ARRAY[PERSON(ID=2,NAME=Peter Smith),PERSON(ID=1,NAME=Fred Smith),PERSON(ID=3,NAME=Mary Smith)]|Lee Smith |
|ARRAY[PERSON(ID=2,NAME=Peter Smith),PERSON(ID=1,NAME=Fred Smith),PERSON(ID=3,NAME=Mary Smith)]|Bill Smith|
|ARRAY[PERSON(ID=2,NAME=Peter Smith),PERSON(ID=1,NAME=Fred Smith)]                        |Mary Smith|
|ARRAY[PERSON(ID=2,NAME=Peter Smith)]                                                     |Fred Smith|
|-----------------------------------------------------------------------------------------|----------|
SQL> alter table person add primary key(name)
SQL> alter table person drop id
SQL> create role ps
SQL> grant ps to "MALCOLM1\Malcolm"
SQL>
```

Figure 3. This shows the commands needed in our implementation to create a new database containing the example graph data, some simple graph-oriented queries, and some steps to develop the model and make it available to the network

```
MATCH ({name:'Peter Smith'}) [()-[:Child]->()]+
(x) RETURN x.name
```

will yield a list of the descendants of Peter Smith.

Without using RETURN or any dependent statements, the result of a MATCH statement is the list of bindings. The above example has two columns, one for each of the unbound identifiers p and x, but p will be an array with an element for each iteration of the pattern.

The results are shown in Figure 3, which also shows all of the statements needed in our implementation to build and display this small example, including two lines for replacing the default primary key ID. A feature of the implementation described in this paper is the lack of structural clutter.

In sections B and C, we continue this small example with two further steps, to display the contents as a graph, and to show how the relational database directly supports object-oriented application programming for such graphical data.

### B. Graph versus Relation

The nodes and edges contained in the database combine to form a set of disjoint graphs that is initially empty. Adding a node to the database adds a new entry to this set. When an edge is added, either the two endpoints are in the same graph, or else the edge will connect two previously disjoint graphs. If each graph in the set is identified by a representative node (such as the one with the lowest uid) and maintains a list of the nodes and edges it contains, it is easy to manage the set of graphs as data is added to the database.

If an edge is removed, the graph containing it might now be in at most two pieces: the simplest algorithm removes it from the set and adds its nodes and edges back in.

It is helpful if the RDBMS is extended to provide a graphical display as in Figure 2 above. In our work the RDBMS provides a simple HTTP service, so that once the database has given appropriate authorization an ordinary web access will display the graph in a browser. Selection of a node with the mouse displays its properties.

The database with its added graph information can be used directly in ordinary database application processing, with the advantage of being able to perform graph-oriented querying and graph-oriented stored procedures. The normal processing of the database engine naturally enforces the type requirements of the model, and also enforces any constraints specified in graph-oriented metadata. The nodes and edges are rows in ordinary tables that can be accessed and refined using normal SQL statements. In particular, using the usual dotted syntax, properties can be SET and updated, and can be removed by being set to NULL.

### C. Database Design by Example

From the above description of the CREATE statement, we can see that this mechanism allows first versions of types and instances to be developed together, with minimal schema indications. The MATCH statement allows extension of the design by retrieving instances and creating related nodes and edges.

If example nodes and edges are created, the DBMS creates suitable node and edge types, modifying these if additional properties receive values in later examples.

Since transactions are supported, tentative examples can be explored and rolled back or committed. Alter statements can change names, enhance property types and modify subtype relationships, and the SQL Cast function can be used to parse the string representation of a structure value. The usual restrict/cascade actions are available, and node and edge types can have additional constraints, triggers, and methods. As each node and edge type has an associated base table in the database, the result of this process is a relational database that is immediately usable.

As the TGM is developed and merged with other graphical data, conflicts will be detected and diagnostics will help to identify any obstacles to integrating a new part of the model, so that the model as developed to that point can be refined. The SQL ALTER TABLE and ALTER TYPE statements, together with a metadata syntax, allow major changes to the model to be performed automatically, e.g., to enforce expectations on the data.

It is important that all such major changes, indeed all cascades and trigger side effects, are validated as part of the transaction commit process, so that the database is not left in an inconsistent state as a result of a mistake or security exception. An example of such a cascade occurs where a graph has been created using the server's autokey mechanism for primary keys, and the analyst has identified a more suitable numeric or string-valued key. A single ALTER TABLE statement can install this as the new primary key and the change automatically propagates to the edge types that attach to the node type in question. The previous primary key remains as a unique key but can later be dropped without losing any information. Figure 3 shows this process, and its consequences are visible in Figures 2 and 4.

Other restructuring of node types can be performed with the help of the CAST function, which can be used to parse complex types from strings, array and set constructors, and UNNEST. Node and edge manipulations can also be performed by triggers and stored procedures.

The points covered in the above section already go a long way towards an integrated DBMS product that supports the TGM. The resulting TGM implementation inherits aspects such as transacted behavior, constraints, triggers, and stored procedures from the relational mechanisms, since Match and Create statements are implemented as Procedure Statements. The security model in the underlying RDBMS, with its users, roles, and grants of privileges also applies to the base tables and hence to the graphs. Node and edge types emerge as a special kind of structured type. It is thus a relatively simple matter to support view-mediated remote access and object-oriented entity management. Nodes and edges are entities and the same access and Multiple Version Concurrency Control (MVCC) models in our previous work [11] transfer with little trouble into the new features.

As the TGM is developed and merged with other graphical data, conflicts will be detected and diagnostics will help to identify any obstacles to integrating a new part of the model, so that the model as developed to that point can be refined.

It is natural to expect a user interface that displays a graphical version of the property graph. Figure 2 was generated by sending a link (see caption of Figure 2) to our implementation's HTTP service to draw a picture of a portion of a graph starting at a given node. Selection of a node or edge displays the properties of that node and links to redraw the graph starting at another node.

Our database server implementation has for years generated classes for C#, Python or Java applications corresponding to versioned database objects. Here this leads to object-oriented application programming, where node and edge types correspond to classes whose instances are nodes and edges. The Match and Create statements can be used (a) for SQL clients in commands and prepared statements, (b) in the generated C#, Java or Python and the widely used database connection methods ExecuteReader and ExecuteNonQuery, or (c) in JavaScript posted to the web service interface of the database server. In Figure 4 we show a portion of a C# application program to display the descendants of Peter Smith in the little example graph database discussed above.

The normal processing of the database engine naturally enforces the type requirements of the model, and also enforces a range of constraints specified in graph-oriented metadata. The nodes and edges are rows in ordinary tables that can be accessed and refined using normal SQL statements. In particular, using the usual dotted syntax, properties can be SET and updated, and can be removed by being set to NULL.

```csharp
/// <summary>
/// EdgeType CHILDOF from Database ps, Role PS
/// PrimaryKey(ID)
/// ForeignKey, CascadeUpdate(PARENT) PERSON
/// ForeignKey, CascadeUpdate(CHILD) PERSON
/// </summary>
[EdgeType(164, 533)]
5 references
public class CHILDOF : Versioned
{
    [Identity]
    [Field(PyrrhoDbType.Integer)]
    [AutoKey]
    public Int64? ID;
    [Leaving]
    [Field(PyrrhoDbType.String)]
    public String? PARENT;
    [Arriving]
    [Field(PyrrhoDbType.String)]
    public String? CHILD;
    0 references
    public PERSON? PARENTis => conn?.FindOne<PERSON>(("NAME", PARENT));
    1 reference
    public PERSON? CHILDis => conn?.FindOne<PERSON>(("NAME", CHILD));
}
0 references
public class Demo
{
    static PyrrhoConnect? conn = null;
    2 references
    static List<PERSON> Descendants(PERSON p)
    {
        var ds = new List<PERSON>();
        if (p.ofPARENTs is CHILDOF[] ca)
            foreach (var c in ca)
                if (c.CHILDis is PERSON d)
                {
                    ds.Add(d);
                    ds.AddRange(Descendants(d));
                }
        return ds;
    }
    0 references
    static void Main()
    {
        conn = new PyrrhoConnect("Files=ps;Role=PS");
        conn.Open();
        try
        {
            // Get a list of all descendants of Pete Smith
            var pa = conn.FindWith<PERSON>(("NAME","Peter Smith"));
            if (pa.Length == 1)
                foreach (var c in Descendants(pa[0]))
                    Console.WriteLine(c.NAME);
        }
        catch (Exception ex)
```

Figure 4. A portion of a C# application program to find the descendants of Peter Smith in the example database above

IV. AN EXAMPLE

Examples for a graph structure usually choose social networks. We want to show that the TGM is equally suitable for Enterprise Resource Planning (ERP) and other business systems. As a non-trivial example, we have chosen a commercial enterprise which buys parts and products, resells the purchased products or assembles products from purchased parts and sells these value-added products. It does not develop and construct products from raw material but add some value to parts or assembles some products to form systems.

The data model shown above in Figure 1 is suitable for a customer-supplier ordering system and comprises 3 company divisions or departments: sales (green), stock (blue), and procurement (red). These are framed in Figure 1(a) with a green dashed line for sales data, with blue for stock data, and red for procurement or purchase. The graph schema is visualized using UML notation which allows specifying the cardinality of the edges. The correspondence between the Typed Graph Schema (TGS) elements and the UML is shown in Table I.

TABLE I. TGS CORRESPONDENCE WITH UML NOTATION

| TGS | UML |
|---|---|
| $n \in N_S$ | class |
| $e \in E_S$ | association |
| $t = (l, d) \in T$ | $l$ = name of $n$ resp. $e$; $d$ = type of $n$ resp. $e$ |
| $\varrho$ ($e$) | all ends of $e$ |
| $\tau_e(n)$ | (min,max)-cardinality of $e$ at $n$ |
| $C$ | constraints in [ ] or { } |

The sales division needs to manage customer data and process the customer's orders. It consists of Customer nodes with properties CustNo, Name and Address. The Name and Address might as well be structured data types for first- and last name resp. street, ZIP code, and city. The CustOrder node mainly comprises OrdNo, the (redundant) CustNo, order date Date and the order total Sum in Euros. The CustOrder contains 1 to many order detail lines of OrderPos which consist at least of the order quantity as property. The order quantity itself is suppressed in the UML diagram to avoid overloading the picture. According to the semantics of the TGM the edge arrows signify the reading direction of the edge type. In the case of "belongs_to" the reading direction is from OrderPos to CustOrder.

All other necessary properties for an order line (e. g. partNo, PartName, UnitProce) could be determined by following the edges of the model to the Part, Stock, and CustOrder node. In Figure 1 (a) only the nodes Customer and CustOrder are showing exemplified properties. More properties are maintained in a real situation, e. g. planned delivery, shipping date, etc for a customer order. The same applies to all other nodes, e. g. unit and quantity discount for parts.

The procurement division is responsible for maintaining the supplier data and ordering of parts and products from them. It mirrors the sales model structurally and comprises supplier, the purchases (SupplOrd, PurchPos) and the supplier catalogue. Purchase- and Sales division have connections to the stock management.

Finally, the stock division comprises master data management and stock management. Master data management includes structural information about the parts in the form a Bill Of Materials (BOM). Stock management deals with adding parts to the stock and releaseing them from stock, The central node of the stock model is the Part node who distinguishes between purchased parts (PurchasedParts) and in-house products (InHouseProduct) modelled as subtypes of Part. We have a BOM structurally represented as a recursive edge "part_of" on the part nodes. The BOM forms a tree structure with the product at the top. The product is made up recursively of components (composed parts) and finally of single parts. The stock itself is represented as a node with properties like number of parts, reservations, and commissions. A stock node is linked to a part and a storage location. This allows knowing exactly which part is located at a certain location in the warehouse.

Figure 1 (b) gives a high level view on the scenario. Such kinds of abstractions are important for complex graphs in order to keep the model manageable. CASE tools that support zoom-in and zoom-out functions would be beneficial to assist the graph modelling.

The syntax of the above presented example ERP model will be presented in the following subsection. Multiline statements are enclosed in square brackets.

### A. Syntax of the ERP example

First we start with the sales graph (green schema), followed by the supplier (red schema) and stock division (blue schema), and finally the three divisions are linked by the edge types "serves", "supplies", "canSupply", "orders", and "from".

The green schema is illustrated in Figure 5 below, and the declarations:

```
// sales division
[CREATE
(a:Customer {CustNo:1001, Name:'Adam', Address:'122, Nutley Terrace, London, ST 7UR, GB'} ), // Customer
(b:Customer {CustNo:1002, Name:'Brian', Address:'45, Belsize Square, London, ST 7UR, GB'} ),
 // …
(f:Customer {CustNo:1006, Name:'Eddy', Address:'72, Ibrox Street, Glasgow, G51 1AA, UK'} ), // customer without order
 (o1:CustOrder    {OrdNo:2001,    CustNo:1001, Datum:DATE'2023-03-22', SummE:211.00} ),    // CustOrder
(o2:CustOrder    {OrdNo:2002,    CustNo:1002, Datum:DATE'2023-03-22', SummE:24.00} ),
// …
(o8:CustOrder    {OrdNo:2008,    CustNo:1002, Datum:DATE'2023-04-24', SummE:808.00} ),
 (op1:OrderPos {Quantity:4, Unit:'piece'} ),    // OrdPos
(op2:OrderPos {Quantity:4, Unit:'litre'} ),
// …
(op18:OrderPos {Quantity:10, Unit:'piece'} ),
(a)<-[:ORDERED_BY]-(o1),    // each order was ordered by exactly 1 customer
(a)<-[:ORDERED_BY]-(o6),
(a)<-[:ORDERED_BY]-(o7),
(b)<-[:ORDERED_BY]-(o2),
//…
(o1)<-[:BELONGS_TO]-(op1),    // each orderPos belongs to exactly 1 order
(o2)<-[:BELONGS_TO]-(op2),

// …
(o8)<-[:BELONGS_TO]-(op9), // and an order has at least 1 orderPos
(o8)<-[:BELONGS_TO]-(op10),
(o1)<-[:BELONGS_TO]-(op11),
// …
   (o8)<-[:BELONGS_TO]-(op18)]
```

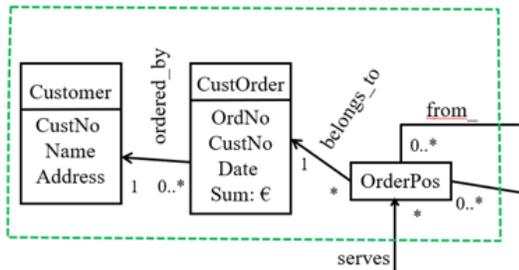

Figure 5. The customer section of the database (from Figure 1)

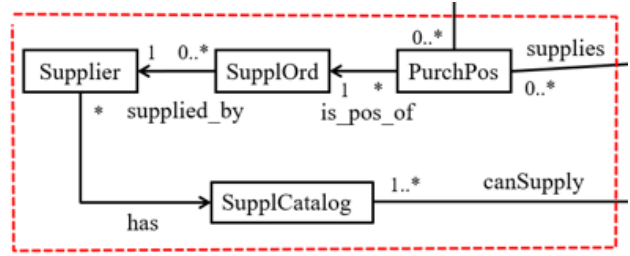

Figure 6. The Supplier part of the example database from Figure 1

We continue with the supplier division, illustrated in Figure 6 below. Here are sample declarations for this section:

Next, the Stock part, shown in Figure 7 below. Here are sample declarations:

```
// supplier division
[ CREATE
(a:Supplier {SupplNo:101, Name:'Rawside Furniture',
Address:'58 City Rd, London , EC1Y 2AL, UK'} ),
(b:Supplier {SupplNo:102, Name:'Andreas Stihl Ltd',
Address:'Stihl House Stanhope Road, GU 15 3 YT,
Camberley Surrey, GB'} ),
// …
// SupplOrd
(o1:SupplOrd      {OrdNo:2001,      SupplNo:101,
Datum:DATE'2023-01-11', "Sum€":260.00} ),
(o2:SupplOrd      {OrdNo:2002,      SupplNo:102,
Datum:DATE'2023-02-22', "Sum€":2405.00} ),
// …
// OrdPos purchase details
(op1:PurchOrd {PosNo:1, Quantity:4, Unit:'piece'} ),
(op2:PurchOrd {PosNo:1, Quantity:4, Unit:'litre'} ),
// …
// (Supplier)<-[:SUPPLIED_BY]-(SupplOrd)
(a)<-[:SUPPLIED_BY]-(o1),  // each order was ordered
by exactly 1 Supplier
(a)<-[:SUPPLIED_BY]-(o4),
// …
// (SupplOrd)<-[:IS_POS_OF]-(OrdPos)
(o1)<-[:IS_POS_OF]-(op1),  // each PurchPos belongs
to exactly 1 order
(o2)<-[:IS_POS_OF]-(op2),
// …
(o1)<-[:IS_POS_OF]-(op7),  // and an order has at
least 1 PurchPos
(o1)<-[:IS_POS_OF]-(op8),
(o1)<-[:IS_POS_OF]-(op9),
//…
// SupplCatalog
(sc11:SupplCatalog     {SupplNo:101,   SPartNo:'sp1',
description:'Hammer handle, Wood (ash), Weight:100
g', unit:'piece', unitPrice:2.00}), //P15
(sc12:SupplCatalog       {SupplNo:101,SPartNo:'sp2',
description:'Tabletop, Wood (oak), Color:brown,
Size:80w x120l cm', unit:'piece', unitPrice:40.00}),
//P16
// …
(sc46:SupplCatalog     {SupplNo:104,   SPartNo:'sp6',
description:'Shelf spruce, Color: white, Weight:6 kg,
Size:60w x180h cm', unit:'piece', unitPrice:20}),
// (Supplier)-[:HAS]->(SupplCatalog)
(a)-[:HAS]->(sc11), (a)-[:HAS]->(sc12), (a)-[:HAS]-
>(sc13), (a)-[:HAS]->(sc14), (a)-[:HAS]->(sc15), (a)-
[:HAS]->(sc16),
    (b)-[:HAS]->(sc21),   (b)-[:HAS]->(sc22),   (b)-
[:HAS]->(sc23),   (b)-[:HAS]->(sc24),   (b)-[:HAS]-
>(sc25)]
```

```
// stock division
// create Part types
create type Part as (PartID char ,Designation char,
Color char, Weight char, Size char) nodetype
// PurchasedPart
create    type    PurchasedPart    under    Part    as
(PreferredSupplNo int,  sumOrderedThisYear currency,
discountPrice currency)
// InHouseProduct
create    type    InHouseProduct    under    Part    as
(ProductionPlan    char,    producedThisYear    int,
manufacturingCosts currency)
[CREATE
(a1:Location {LocationNo:10011, Aisle:1, Shelf:'left
A', Rack: 'A1'} ),  // Location
(a2:Location {LocationNo:10012, Aisle:1, Shelf:'left
A', Rack: 'A2'} ),
// …
(l:Location {LocationNo:10111, Aisle:2, Shelf:'left
A', Rack: 'A1'} ),  // Location without parts
//Part will be filled implicitly
// PurchasedPart
(p1:PurchasedPart                        {PartID:'P01'
,Designation:'Wallplug',Material:'Fiber',
Color:'grey',     Weight:'6     g',     Size:'12    cm',
PreferredSupplNo:103,         sumOrderedThisYear:2000,
discountPrice:'0.04 €'   }),  //p1 Wallplug
```

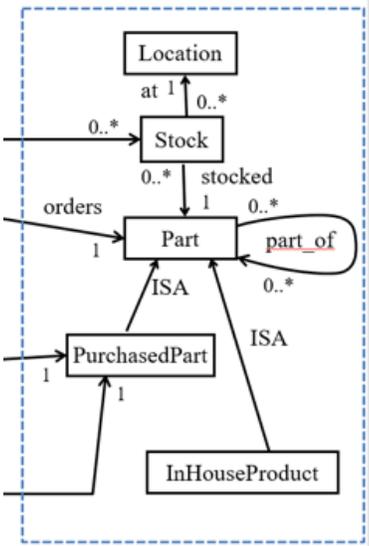

Figure 7. The Stock part of the example database from Figure 1

```
(p5:PurchasedPart   {PartID:'P05'   ,Designation:'Metal
nail', Material:'Metal', Color:'grey', Weight:'2 g',
Size:'A 50 x2.2 mm',
PreferredSupplNo:102,      sumOrderedThisYear:10000,
discountPrice:'0.005 €'}),   //p5 Metal nail
// …
(p30:PurchasedPart                        {PartID:'P30'
,Designation:'Degreasing liquid', Material:'benzine',
Color:'clear', Weight:'100 g', Size:'100 ml bottle' ,
    PreferredSupplNo:101,   sumOrderedThisYear:150,
discountPrice:'1.80 €'}), //p30 Degreasing liquid
// InHouseProduct
(p2:InHouseProduct {PartID:'P02'  ,Designation:'Power
plug', Color:'white', Weight:'30 g', Size:'dia 5 cm
',
ProductionPlan:'P02 Power plug',
producedThisYear:1000, manufacturingCosts:'2.50 €'}),
(p3:InHouseProduct                        {PartID:'P03'
,Designation:'Hammer',         Material:'Compound
material',Color:'blue', Weight:'1,1 kg', Size:'35 cm
long',
ProductionPlan:'P03  Hammer',   producedThisYear:100,
manufacturingCosts:'2.50 €'}),
// …
(p28:InHouseProduct                       {PartID:'P28'
,Designation:'Tableleg',
Material:'Metal',Color:'Silver',          Weight:'1
kg',Size:'80w x120l cm',
ProductionPlan:'P28 Tableleg',  producedThisYear:160,
manufacturingCosts:'7.00 €'}),
```

```
// Stock
(s1:Stock      {PartID:'P02',     LocationNo:10011,
available:55,                      commissioned:20,
reserved_until:DATE'2023-09-22'} ),
(s2:Stock      {PartID:'P11',     LocationNo:10012,
available:500,                    commissioned:100,
reserved_until:DATE'2023-10-12'} ),
// …
(s34:Stock     {PartID:'P30',     LocationNo:10101,
available:30,                       commissioned:5,
reserved_until:DATE'2024-09-21'} ),
 //BOM
(p2)<-[:IS_Part_OF {no_of_components:1}]-(p11),
(p2)<-[:IS_Part_OF    {no_of_components:2}]-(p12)<-
[:IS_Part_OF {no_of_components:1}]-(p13),
(p3)<-[:IS_Part_OF {no_of_components:1}]-(p14),
// …
(p26)<-[:IS_Part_OF {no_of_components:1}]-(p23),
// Links: Parts<-Stock->Location
(p1)<-[:stocked]-(s33)-[:at]->(i3),
(p2)<-[:stocked]-(s1)-[:at]->(a1),
// …
(p30)<-[:stocked]-(s34)-[:at]->(k)]
```

Table II summarizes the schema objects (node and edge types) of the ERP graph schema and Figure 8 shows part of the resulting graph view of the database.

TABLE II. NODE AND EDGE TYPES IN AN EXAMPLE DATABASE (RELATIONAL DESCRIPTION)

| Type name | Informal Description | SuperType |
|---|---|---|
| Customer | (CustNo, Name, Address) | |
| CustOrder | (CustNo, Datum, OrdNo, Summ€) | |
| OrderPos | (Id, Quantity, Unit) | |
| Location | (LocationNo, Reihe, Shelf, Rack) | |
| PurchasePart | (PartID, Designation, Material, Color, Weight, Size) | Part |
| InHouseProduct | (PartID, Designation, Material, Color, Weight, Size) | Part |
| Stock | (PartID, LocationNo, Available, Commissioned, Reserved_Until) | |
| Supplier | (SupplNo, Name, Address) | |
| SupplOrd | (OrdNo, SupplNo, Datum, Sum€) | |
| PurchPos | (PosNo, Quantity, Unit) | |
| SupplCatalog | (SupplNo, SPartNo, Desription, Weight, Unit, unitPrice) | |

| Type name | Leaving | Arriving | Other properties |
|---|---|---|---|
| Ordered_by | CustOrder | Customer | |
| Belongs_to | OrderPos | CustOrder | |
| Is_Part_Of | Part | Part | No_of_components |
| Stocked | Stocked | Part | |
| At | Part | Location | |
| Supplied_by | SupplOrd | Supplier | |
| Is_Pos_of | PurchPos | SupplOrd | |
| Has | Sypplier | SupplCatalog | |
| Orders | OrderPos | Part | |
| From_ | OrderPos | Stock | |
| Supplied | PurchPos | ParchasePart | |
| Can_Spply | SupplCatalog | PurchasePart | |
| Serves | PurchPos | OrderPos | |

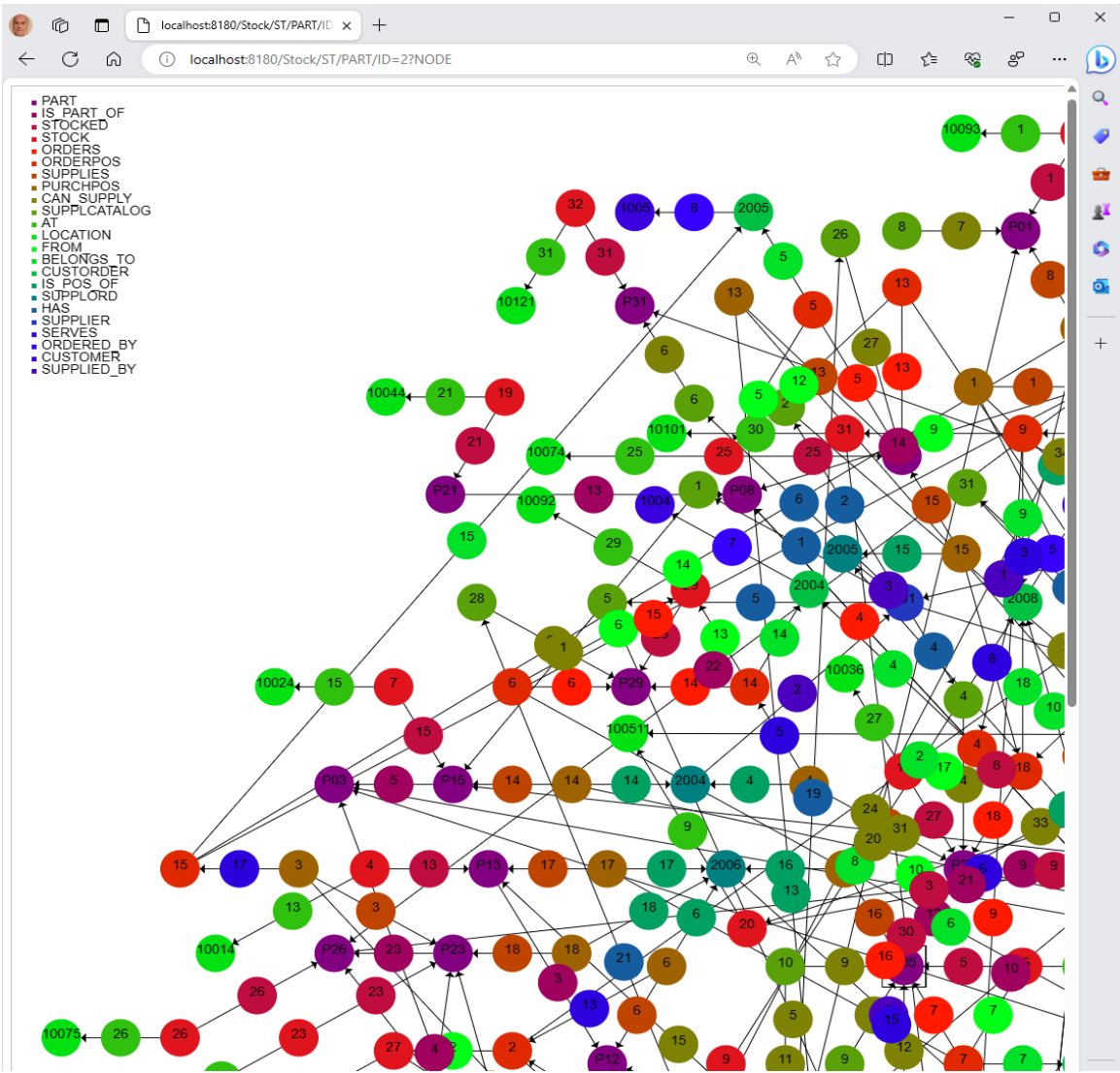

Figure 8. A part of the ERP example graph, after changes to primary keys similar to Figure 2 and 3 (e.g., PART now has key PartID).

V. CONCLUSIONS

The purpose of this paper was to report some progress in our Typed Graph Modeling workstream. The work is available on Github [11] for free download and use and is not covered by any patent or other restrictions. The main challenges, as expected, were related to the implementation of the MATCH algorithm for repeating patterns, and the solution found is an elegant one involving continuations and documented in the Pyrrho blog [12] and in [11]. We plan to add further facilities for altering the types of graph properties, and to track development of the forthcoming GQL standard.

Unsurprisingly, the performance of our implementation is modest for complex statements when the database becomes large. Simple CREATE and MATCH statements like those found in benchmarks are processed at over 2500 per second. The implementation will no double benefit from a review of this aspect.

The current "alpha" state of the software implements all of the above ideas. The test suite includes simple cases that demonstrate the integration of the relational and typed graph model concepts in Pyrrho DBMS. The implementation is backward compatible with previous versions of Pyrrho DBMS, so legacy databases can immediately use these new capabilities. Pyrrho DBMS is free standing and works directly with the operating system (Windows, Linux, or MacOS), and clients interact with the server using TCP/IP or HTTP.

It is our hope that other DBMS developers will also adopt GQL in new versions of their DBMS.

REFERENCES

[1] F. Laux and M. Crowe, "Typed Graph Models and Relational Database Technology", DBKDA 2023: The Fifteenth International Conference on Advances in Databases, Knowledge, and Data Applications, IARIA, March 2023, pp. 33-37, ISSN: 2308-4332, ISBN: 978-1-68558-056-8

[2] ISO 9075-16 Property Graph Queries (SQL/PGQ), International Standards Organisation (2023).

[3] F. Laux and M. Crowe, "Information Integration using the Typed Graph Model", DBKDA 2021: The Thirteenth International Conference on Advances in Databases, Knowledge, and Data Applications, IARIA, May 2021, pp. 7-14, ISSN: 2308-4332, ISBN: 978-1-61208-857-0

[4] E. J. Naiburg and R. A. Maksimschuk, UML for database design. Addison-Wesley Professional, 2001

[5] R. De Virgilio, A. Maccioni, R. Torloner, "Model-Driven Design of Graph Databases", in Yue, E. et al (eds) Conceptual Modeling, 33rd International Conference (ER 2014), Springer, Oct 2014, pp. 172-185, ISSN: 0302-9743 ISBN: 978-3-319-12205-2

[6] F. Laux, "The Typed Graph Model", DBKDA 2020 : The Twelfth International Conference on Advances in Databases, Knowledge, and Data Applications, IARIA, Sept 2020, pp. 13-19, ISSN: 2308-4332, ISBN: 978-1-61208-790-0

[7] M. Crowe and F. Laux, "Database Technology Evolution", IARIA International Journal on Advanced is Software, vol. 15, numbers 3 and 4, 2022, pp. 224-234, ISSN: 1942-2628

[8] S. Shah et al., The PostgreSQL Data Computing Platform (PgDCP) (Online), Available from: https://github.com/netspective-studios/PgDCP [retrieved: Aug 2023]

[9] N. Francis, A. Gheerbrant, P. Guagliardo, L. Leonid, V. Marsault, et al., A Researcher's Digest of GQL. 26th International Conference on Database Theory (ICDT 2023), Mar 2023, Ioannina, Greece. doi:10.4230/LIPIcs.ICDT.2023.1. https://hal.science/hal-04094449 [retrieved: Aug 2023]

[10] M. Laiho and F. Laux, An Introduction to Neo4j Graph Database, DBTechNet.org preprint [retrieved: Aug 2023].

[11] M. Crowe, PyrrhoV7alpha, https://github.com/MalcolmCrowe/ShareableDataStructures [retrieved: Nov, 2023]

[12] M. Crowe, PyrrhoDBMS http://pyrrhodb.com [retrieved Nov, 2023]